%% file: MICCAI2026-main_conference_paper_template.tex
\documentclass[runningheads]{llncs}
\usepackage[T1]{fontenc}
\usepackage{graphicx}
\usepackage{wrapfig}
\usepackage{amsmath}
\usepackage{amssymb}
\usepackage{booktabs}
\usepackage{hyperref}
\usepackage{url}
\usepackage{float}
\usepackage{multirow}
\usepackage{xcolor}
\usepackage{colortbl}

\definecolor{lightblue}{rgb}{0.95,0.97,1.0} 
\definecolor{highlight}{rgb}{0.9,0.95,1.0} 

\input{math_commands.tex}

\begin{document}

\title{Graph-Based Multi-Modal Light-weight Network for Adaptive Brain Tumor Segmentation}

\titlerunning{GMLN-BTS}

\author{Guohao Huo\inst{1}, Ruiting Dai\inst{1}, Zitong Wang\inst{1}, Junxin Kong\inst{1} \and
Hao Tang\inst{2}\thanks{Corresponding author.}}
\authorrunning{G. Huo et al.}
\institute{University of Electronic Science and Technology of China \and
Peking University\\
\email{gh.huo513@gmail.com, haotang@pku.edu.cn}}
\maketitle

\begin{abstract}
Multi-modal brain tumor segmentation remains challenging for practical deployment due to the high computational costs of mainstream models. In this work, we propose GMLN-BTS, a Graph-based Multi-modal interaction Lightweight Network for brain tumor segmentation. Our architecture achieves high-precision, resource-efficient segmentation through three key components. First, a Modality-Aware Adaptive Encoder (M2AE) facilitates efficient multi-scale semantic extraction. Second, a Graph-based Multi-Modal Collaborative Interaction Module (G2MCIM) leverages graph structures to model complementary cross-modal relationships. Finally, a Voxel Refinement UpSampling Module (VRUM) integrates linear interpolation with multi-scale transposed convolutions to suppress artifacts and preserve boundary details. Experimental results on BraTS 2017, 2019, and 2021 benchmarks demonstrate that GMLN-BTS achieves state-of-the-art performance among lightweight models. With only 4.58M parameters, our method reduces parameter count by 98\% compared to mainstream 3D Transformers while significantly outperforming existing compact approaches.

\end{abstract}

\section{Introduction}
Accurate brain tumor segmentation is essential for automated lesion localization and clinical diagnosis \cite{BraTS}. Although multi-modal MRI remains the gold standard for brain tumor assessment \cite{GoldLevel}, current segmentation models often suffer from excessive parameter counts. These heavy architectures hinder efficient deployment in resource-constrained clinical environments \cite{MRI2,MRI3}. Maintaining high segmentation performance while compressing model scale represents a critical challenge for lightweight design. To address this, we propose GMLN-BTS, a Graph-based Multi-modal Lightweight Network for adaptive brain tumor segmentation. By leveraging an efficient cross-modal feature fusion mechanism, our framework significantly reduces computational overhead. Simultaneously, it enhances model adaptability to diverse imaging conditions and improves segmentation precision.

Effective brain tumor segmentation relies on the synergy among multiple MRI modalities, including FLAIR, T1ce, T1, and T2 \cite{BraTS}. These sequences provide complementary structural and pathological contrast information \cite{BTSDNN}. Modeling inter-modal dependencies is crucial because different modalities exhibit distinct sensitivities to specific tumor sub-regions. For instance, FLAIR highlights edema while T1ce emphasizes necrotic cores. We propose the Graph-based Multi-Modal Collaborative Interaction Module (G2MCIM) to address this. This module leverages graph node interactions and edge relationship modeling to explicitly capture and enhance cross-modal complementary features. Consequently, G2MCIM facilitates more robust representation learning for complex tumor structures.

Accurate 3D tumor boundary reconstruction requires effective upsampling within the decoder \cite{V-Net}. Conventional methods like linear interpolation offer stability but suffer from low-frequency blurring. Conversely, transposed convolutions recover high-frequency details yet frequently introduce checkerboard artifacts \cite{InterTranspose}. To leverage their respective advantages, we propose the Voxel Refinement UpSampling Module (VRUM). By synergistically integrating interpolative stability with the detail-recovery capability of transposed convolutions, VRUM effectively suppresses artifacts while preserving high-frequency boundary details. This collaborative mechanism ensures superior segmentation precision across complex anatomical structures.

In summary, our contributions are as follows:

\begin{enumerate}
\item We propose a Modality-Aware Adaptive Encoder (M2AE) that leverages 3D Inception modules for multi-scale feature extraction. This component utilizes Group Normalization to ensure distributional stability and achieves superior representation capability.
\item We introduce a Graph-based Multimodal Collaborative Interaction Module (G2MCIM). By constructing inter-modal relationship graphs, this module adaptively learns cross-modal dependencies to effectively model the differential sensitivity of MRI modalities to tumor sub-regions.
\item We design a Voxel Refinement UpSampling Module (VRUM) that synergistically fuses linear interpolation with multi-scale transposed convolutions. VRUM preserves high-frequency boundary details while effectively suppressing artifacts to enable precise segmentation.
\item Our GMLN-BTS achieves state-of-the-art performance on BraTS 2017, 2019, and 2021 datasets. With only 4.58M parameters (a 98\% reduction compared to nnFormer), it significantly outperforms existing lightweight models and approaches the accuracy of heavy-duty 3D Transformers.
\end{enumerate}

\section{Related Work}

\noindent\textbf{Multimodal Brain Tumor Segmentation.}
The core challenge of deep learning in multi-modal brain tumor segmentation lies in effectively modeling inter-modal complementarity. Early CNN-based methods, such as 3D U-Net \cite{3DUNet}, primarily rely on channel concatenation for shallow fusion. While capturing local features, these approaches struggle to explicitly model cross-modal interactions. To enhance global modeling, Transformer-based architectures like TMFormer \cite{TMFormer} introduce self-attention mechanisms to learn inter-modal dependencies. However, their quadratic complexity limits practical clinical utility. Recently, State Space Models (SSMs) like Mamba \cite{mamba} offer a path for efficient sequence modeling with linear complexity. Nevertheless, their cross-modal fusion mechanisms remain largely unexplored. Although hybrid models such as mmFormer \cite{mmformer} and MedSegMamba \cite{medsegmamba} attempt to integrate local and global advantages, they often suffer from coarse-grained interactions and computational redundancy. These developments suggest that designing specialized, lightweight cross-modal interaction mechanisms is critical for practical multi-modal segmentation.

\noindent\textbf{Multimodal Interaction of Brain Tumor Characteristic Information.}
Current multi-modal interaction mechanisms widely employ attention weighting, Transformers, or State Space Models (SSMs) to fuse complementary information. While demonstrating strong representation capabilities, attention-based methods \cite{CMAF-Net,diffbts} achieve adaptive fusion through noise suppression and context modeling at the cost of high computational overhead. Similarly, Transformer and SSM architectures \cite{MicFormer,LS3M,ACMINet} focus on capturing long-range dependencies. However, their dual-stream interactions or dynamic reordering designs significantly increase parameter counts and inference latency. Despite performance gains, these approaches generally suffer from limited computational efficiency and poor deployment feasibility. Consequently, they fail to meet the lightweight and real-time requirements of clinical scenarios. These limitations underscore the urgent need for more efficient feature interaction mechanisms.

\noindent\textbf{Lightweight Multimodal Brain Tumor Segmentation Model.}
To facilitate deployment in resource-constrained clinical environments, researchers have proposed various lightweight segmentation architectures. One category of methods \cite{LATUP-Net,LIU-Net,LR-Net} utilizes parallel or shifted convolutions to achieve structural efficiency. These approaches focus on multi-scale feature extraction and edge refinement but suffer from limited representation capacity and inconsistent inter-modal semantics. Another class of frameworks \cite{SuperLightNet,segformer3d} compresses model complexity via learnable skip connections or multi-scale attention. However, these methods often employ simple modality concatenation, which weakens the effective fusion of complementary cross-modal information. While these strategies emphasize different structural designs, both fail to establish sufficient modality interaction mechanisms. Consequently, these models struggle to capture fine-grained cross-modal correlations. This deficiency in interaction modeling remains a core limitation of current lightweight medical image segmentation methods.

\section{The Proposed Method}
\subsection{Graph-based Multi-Modal Interaction Lightweight Network for Brain Tumor Segmentation (GMLN-BTS)}

\begin{figure}[t]
    \centering
    \includegraphics[width=1\textwidth]{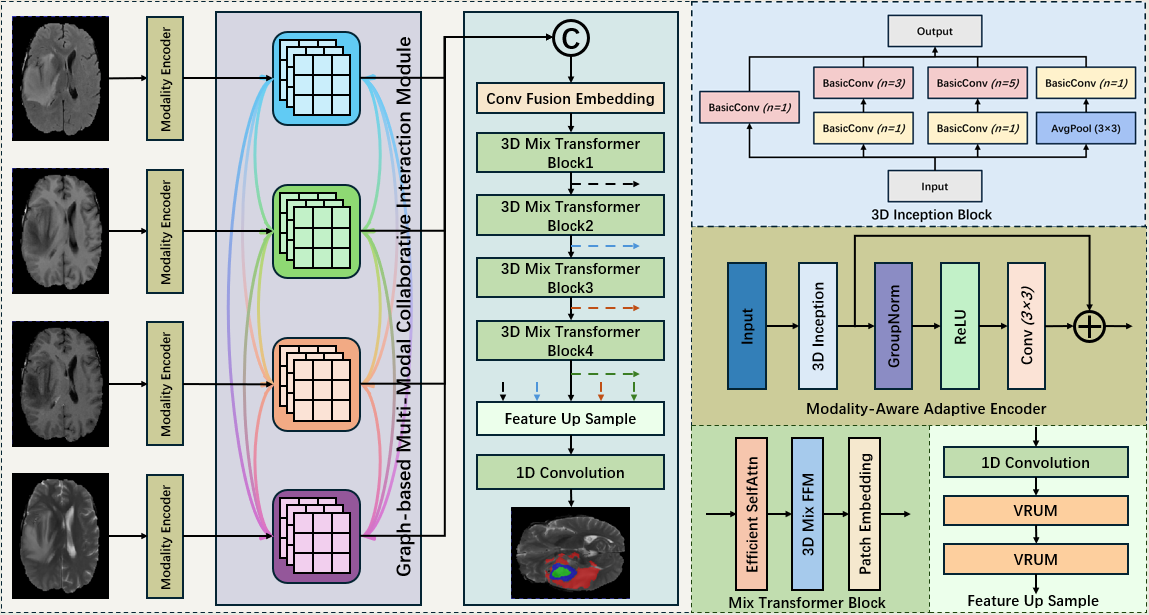}
    \caption{Architectural diagram of the proposed Graph-based Multi-Modal Interaction Lightweight Network for Brain Tumor Segmentation (GMLN-BTS)}
    \label{fig:Model}
\end{figure}

The overall architecture of the proposed GMLN-BTS is illustrated in Figure \ref{fig:Model}. The framework primarily consists of three collaborative components: the Modality-Aware Adaptive Encoder (M2AE) for multi-scale feature encoding, the Graph-based Multi-Modal Collaborative Interaction Module (G2MCIM) for cross-modal dependency modeling, and the Voxel Refinement UpSampling Module (VRUM) for high-fidelity spatial reconstruction. Between the interaction and upsampling stages, a lightweight Transformer block is integrated to capture global context within the fused representations.

\subsection{Modality-Aware Adaptive Encoder (M2AE)}
The proposed M2AE is designed to extract multi-scale semantic features from individual modalities through a 3D Inception Block. Specifically, the block employs parallel convolutional branches with varying kernel sizes to capture diverse receptive fields as follows:
\begin{equation}
\begin{split}
    & \mathrm{Y1}=\mathrm{BasicConv}_{n=1}^{C_{out}/4}(X), \\ & \mathrm{Y2}=\mathrm{BasicConv}_{n=3}^{C_{out}/4}(\mathrm{BasicConv}_{n=1}^{C_r}(X)), \\ & \mathrm{Y3}=\mathrm{BasicConv}_{n=5}^{C_{out}/4}(\mathrm{BasicConv}_{n=1}^{C_r}(X)), \\ & \mathrm{Y4}=\mathrm{BasicConv}_{n=1}^{C_{out}/4}(\mathrm{AvgPool}_{n=3}(X)), \\ & 
    Y_{Modality}=\mathrm{Concat}(Y_1,Y_2,Y_3,Y_4),
\end{split}
\end{equation}
where $Modality \in \{\mathrm{T1, T1ce, T2, FLAIR}\}$. To stabilize feature distributions and enhance representational capacity, we integrate GroupNorm (GN) alongside residual connections and ReLU activations:
\begin{equation}
Z_{\mathrm{Modality}} = \mathrm{Conv3D}(\mathrm{ReLU}(\mathrm{GN}(Y_{\mathrm{Modality}}))) + Y_{\mathrm{Modality}}.
\end{equation}
By setting the output channels $C_2=16$, the M2AE maximizes multi-scale feature richness while maintaining a minimal memory footprint.

\subsection{Graph-based Multi-Modal Collaborative Interaction Module (G2MCIM)}
The proposed G2MCIM leverages the graph structure and its edge relationships to model the interaction of complementary features across modalities for enhanced representation.
First, features encoded by the Modality-Aware Adaptive Encoder across modalities are concatenated to form the output 
$Z_{Out} \in \mathbb{R}^{B \times 4\times C_2 \times D \times H \times W}$:
\begin{align}
    Z_{Out}=\mathrm{Concat}(Z_{T1},Z_{T2},Z_{T1ce},Z_{FLAIR}). 
\end{align} 
To reduce GPU memory consumption, spatial average pooling is applied to compress spatial semantics and extract channel-wise features:
\begin{align}
    V = \frac{1}{D \times H \times W} \sum_{d=1}^{D} \sum_{h=1}^{H} \sum_{w=1}^{W} Z_{Out}.
\end{align}
The output features of the above equations are $V \in \mathbb{R}^{B \times 4 \times C_2}$. Then the cross-modal relationship pairs are constructed:
\begin{equation}
\begin{split}
R = \mathrm{Concat}( & \mathrm{expand}(V, \ \mathrm{dim}=1, \ \mathrm{copies}=4), \mathrm{expand}(V^T, \ \mathrm{dim}=1, \ \mathrm{copies}=4) ).
\end{split}
\end{equation}
The output features of the above equations are $R \in \mathbb{R}^{B \times 4 \times 4 \times 2C_2}$. The modality-specific relation encoding network $\phi_i(\cdot)$ is implemented via bilinear layers as:
\begin{align}
    \phi_i(z)=W_{i2}\cdot\sigma_{\mathrm{LeakyReLU}}(W_{i1}\cdot z+b_{i1})+b_{i2}.
\end{align}
This network is then applied to the output features to generate adaptive relational weights for each modality $i \in \{\mathrm{T1,T1ce,T2,FLAIR}\}$:
\begin{align}
    A_i=\phi_i(R_{:,i,:,:}),
\end{align}
Weight normalization is performed to standardize features and mitigate gradient explosion:
\begin{align}
    S_i = \mathrm{softmax}(A_i) = \frac{\exp(A_{i,j})}{\sum_{k=1}^4 \exp(A_{i,k})}.
\end{align}
The output features of the above equations are $S_i \in \mathbb{R}^{B \times 4 \times C_2}$. Modal features are reshaped as:
\begin{align}
    F=\mathrm{Reshape}(Z_{Out}),
\end{align} 
with $M = D \times H \times W$ and the output features of the above equations are $F \in\mathbb{R}^{B\times4\times C_2\times M}$.
Cross-modal weighted fusion is conducted using relational weights:
\begin{align}
    U_i=\sum_{j=1}^4S_{i,j}\odot F_j,
\end{align} 
where $\odot$ indicates channel-wise multiplication.
The output features of the above equations are $U_i \in\mathbb{R}^{B\times C_2\times M}$. The final output is generated through a residual connection:
\begin{align}
    Y_i=Z_i+\mathrm{Reshape}(U_i,\ [B,C_2,D,H,W]).
\end{align}
The output features of the above equations are $Y_i \in\mathbb{R}^{B\times C_2\times D\times H\times W}$.

\subsection{Voxel Refinement UpSampling Module (VRUM)}
\begin{wrapfigure}{r}{0.5\textwidth}
    \centering
    \vspace{-20pt}
    \includegraphics[width=1\linewidth]{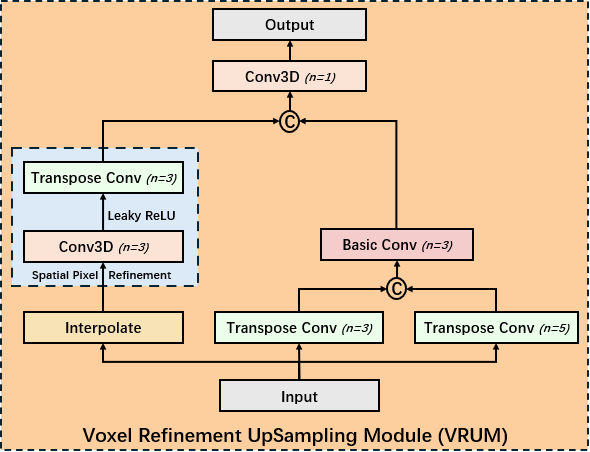}
    \vspace{-20pt}
    \caption{Architecture diagram of the proposed VRUM.}
    \label{fig:VRUM_Structure}
    \vspace{-20pt}
\end{wrapfigure}
The proposed VRUM is designed to reconstruct high-fidelity segmentation boundaries by synergistically integrating linear interpolation and multi-scale transposed convolutions. As illustrated in Figure \ref{fig:VRUM_Structure}, the module adopts a dual-branch architecture to balance structural stability with high-frequency detail enhancement.

\noindent\textbf{Interpolation-based Refinement Branch.} This branch provides a stable feature foundation via trilinear upsampling, followed by a Spatial Pixel Refinement module (comprising 3D convolution and transposed convolution) to recover local sharpness and calibrate spatial semantics:
\begin{equation}
\begin{split}
\mathbf{X}_{\mathrm{interp}} &= \text{TrilinearUpsample}(\mathbf{X}; s=2), \\
\mathbf{X}_{\mathrm{trans}} &= \text{ConvTrans3D}(\mathrm{LeakyReLU}(\mathrm{Conv3D}(\mathbf{X}_{\mathrm{interp}}))).
\end{split}
\end{equation}

\noindent\textbf{Multi-scale Transposed Convolution Branch.} To compensate for the high-frequency information lost during interpolation, this branch employs parallel transposed convolutions with varying kernels ($k=3$ for fine-grained details and $k=5$ for structural coherence) to mitigate checkerboard artifacts while capturing rich textures:
\begin{equation}
\begin{split}
\mathbf{X}_{\mathrm{fuse}} &= \text{Conv3D}(\text{Concat}(\text{ConvTrans3D}_{k=3}(\mathbf{X}), \text{ConvTrans3D}_{k=5}(\mathbf{X}))), \\
\mathbf{X}_{\mathrm{out}} &= \mathrm{ReLU}(\mathrm{BatchNorm}(\mathbf{X}_{\mathrm{fuse}})).
\end{split}
\end{equation}

The final refined output $\mathbf{Y}_{\mathrm{final}}$ is obtained by fusing the features from both branches through concatenation and a $1\times1\times1$ convolution, ensuring a seamless integration of global smoothness and local precision:
\begin{equation}
\mathbf{Y}_{\mathrm{final}} = \mathrm{Conv3D}(\mathrm{Concat}(\mathbf{X}_{\mathrm{trans}}, \mathbf{X}_{\mathrm{out}})).
\end{equation}

\section{Experiments}
\subsection{Datasets and Implementation.} 
We evaluate GMLN-BTS on the BraTS 2017, 2019, and 2021 benchmarks using multi-modal MRI scans (T1, T1ce, T2, FLAIR). The model is implemented in PyTorch on an NVIDIA A800 (80GB) GPU. Following the SegFormer3D protocol, we employ the AdamW optimizer with a base learning rate of $3\times10^{-5}$ (linear warm-up) and PolyLR decay. Training is conducted for 800 epochs with a batch size of 2 using a joint Dice and cross-entropy loss. Results report the mean Dice Similarity Coefficient (DSC) over three independent runs.

\subsection{Quantitative Comparison}
\begin{table}[t] 
    \centering 
    \caption{Quantitative comparison on BraTS datasets.
    \textbf{GMLN-BTS} achieves competitive performance with \textbf{only 4.58M parameters} 
    ($\approx$3$\times$ smaller than SwinUNETR, $\approx$33$\times$ smaller than nnFormer). 
    Best results in \textbf{bold}; \underline{underlined} denotes best among models with $<$5M params.} 
    \label{tab:efficiency_comparison} 
    
    \small 
    \setlength{\tabcolsep}{1.5pt} 
    \renewcommand{\arraystretch}{1.15} 
    
    \resizebox{\textwidth}{!}{%
    \begin{tabular}{l c | c c c c | c c c c | c c c c} 
        \toprule 
        \multirow{2}{*}{\textbf{Method}} & \multirow{2}{*}{\textbf{Params}} & \multicolumn{4}{c|}{\textbf{BraTS 2017}} & \multicolumn{4}{c|}{\textbf{BraTS 2019}} & \multicolumn{4}{c}{\textbf{BraTS 2021}} \\ 
        \cmidrule(r){3-6} \cmidrule(r){7-10} \cmidrule(r){11-14} 
        & (M) & Avg$\uparrow$ & WT & ET & TC & Avg$\uparrow$ & WT & ET & TC & Avg$\uparrow$ & WT & ET & TC \\ 
        \midrule 
        nnFormer~\cite{nnformer}         & 150.50 & \textbf{86.4} & \textbf{91.3} & \textbf{81.8} & \textbf{86.0} & 78.7 & 81.4 & 81.6 & 73.1 & 83.7 & 87.9 & 78.4 & 84.7 \\ 
        SwinUNETR~\cite{SwinUNetr}       & 62.19  & 83.2 & 90.2 & 77.2 & 82.2 & 81.6 & 87.2 & 83.1 & 74.5 & \textbf{89.4} & \textbf{92.7} & \textbf{88.3} & \textbf{90.2} \\ 
        UNETR~\cite{unetr}               & 92.49  & 71.1 & 78.9 & 58.5 & 76.1 & 64.4 & 74.9 & 64.2 & 53.9 & 84.3 & 88.7 & 81.7 & 82.4 \\ 
        \midrule 
        \rowcolor{lightblue}  
        SegFormer3D~\cite{segformer3d}   & 4.51   & 82.1 & 89.9 & 74.2 & 82.2 & 87.9 & 89.6 & 82.4 & 91.8 & 86.0 & 88.9 & 81.4 & 87.8 \\ 
        \rowcolor{lightblue} 
        SuperLightNet~\cite{SuperLightNet}& 2.97  & 77.4 & 84.8 & 66.4 & 80.9 & 84.5 & 81.2 & 80.9 & 91.3 & 86.3 & 88.1 & 84.5 & 86.4 \\ 
        \rowcolor{highlight}  
        \textbf{GMLN-BTS (Ours)}         & \textbf{4.58}$^\star$ & \underline{85.1} & \underline{90.5} & \underline{81.2} & \underline{83.5} & \underline{\textbf{89.4}} & \underline{\textbf{91.3}} & \underline{\textbf{84.9}} & \underline{\textbf{92.0}} & \underline{88.7} & \underline{90.3} & \underline{86.1} & \underline{89.7} \\ 
        \bottomrule 
    \end{tabular} 
    }    
    \vspace{-15pt} 
\end{table}

\noindent\textbf{Lightweight Efficiency.} 
GMLN-BTS establishes a new benchmark for lightweight models ($<5$M parameters). With only \textbf{4.58M parameters}, it consistently outperforms comparable models such as SegFormer3D (4.51M) and SuperLightNet (2.97M). Specifically, GMLN-BTS achieves an average Dice score of \textbf{85.1} on BraTS 2017, significantly surpassing SegFormer3D (82.2) and SuperLightNet (77.4). This performance advantage extends to BraTS 2019 and 2021, yielding scores of \textbf{89.4} and \textbf{88.7}, respectively. These results demonstrate superior feature extraction efficiency under stringent parameter constraints.

\noindent\textbf{Cross-Scale Competitiveness.} 
Compared to heavyweight architectures, GMLN-BTS maintains high competitiveness despite a significantly smaller footprint. Although it is $\sim$33$\times$ smaller than \textbf{nnFormer} (150.50M) and $\sim$13$\times$ smaller than \textbf{SwinUNETR} (62.19M), GMLN-BTS yields superior or comparable accuracy. 
\begin{itemize}
    \item Performance Gain: On BraTS 2019, GMLN-BTS (89.4 Dice) substantially outperforms both nnFormer (78.7) and SwinUNETR (81.6), suggesting a more effective capture of complex tumor boundaries without excessive computational overhead.
    \item Competitive Parity: On the larger BraTS 2021 dataset, GMLN-BTS remains within 1\% of the massive SwinUNETR (88.7 vs. 89.4). Similarly, on BraTS 2017, it remains highly competitive with the state-of-the-art nnFormer (85.1 vs. 86.4) while utilizing only a fraction of the memory and compute.
\end{itemize}

\begin{wraptable}{r}{0.5\textwidth}
    \vspace{-30pt}
    \centering
    \caption{Ablation of GMLN-BTS components (Mean Dice \%).}
    \label{tab:ablation}
    \resizebox{\linewidth}{!}{%
    \begin{tabular}{c|c|c|c|c}
        \hline
        Method & G2MCIM & M2AE & VRUM & Mean Dice \\
        \hline
        Baseline & - & - & - & 81.9 \\
        +G2MCIM & \checkmark & - & - & 84.2 \\
        +M2AE & \checkmark & \checkmark & - & 84.7 \\
        \hline
        \textbf{Full Model} & \textbf{\checkmark} & \textbf{\checkmark} & \textbf{\checkmark} & \textbf{85.1} \\
        \hline
    \end{tabular}
    }
    \vspace{-26pt}
\end{wraptable}

\begin{figure}[t]
    \centering
    \includegraphics[width=0.8\textwidth]{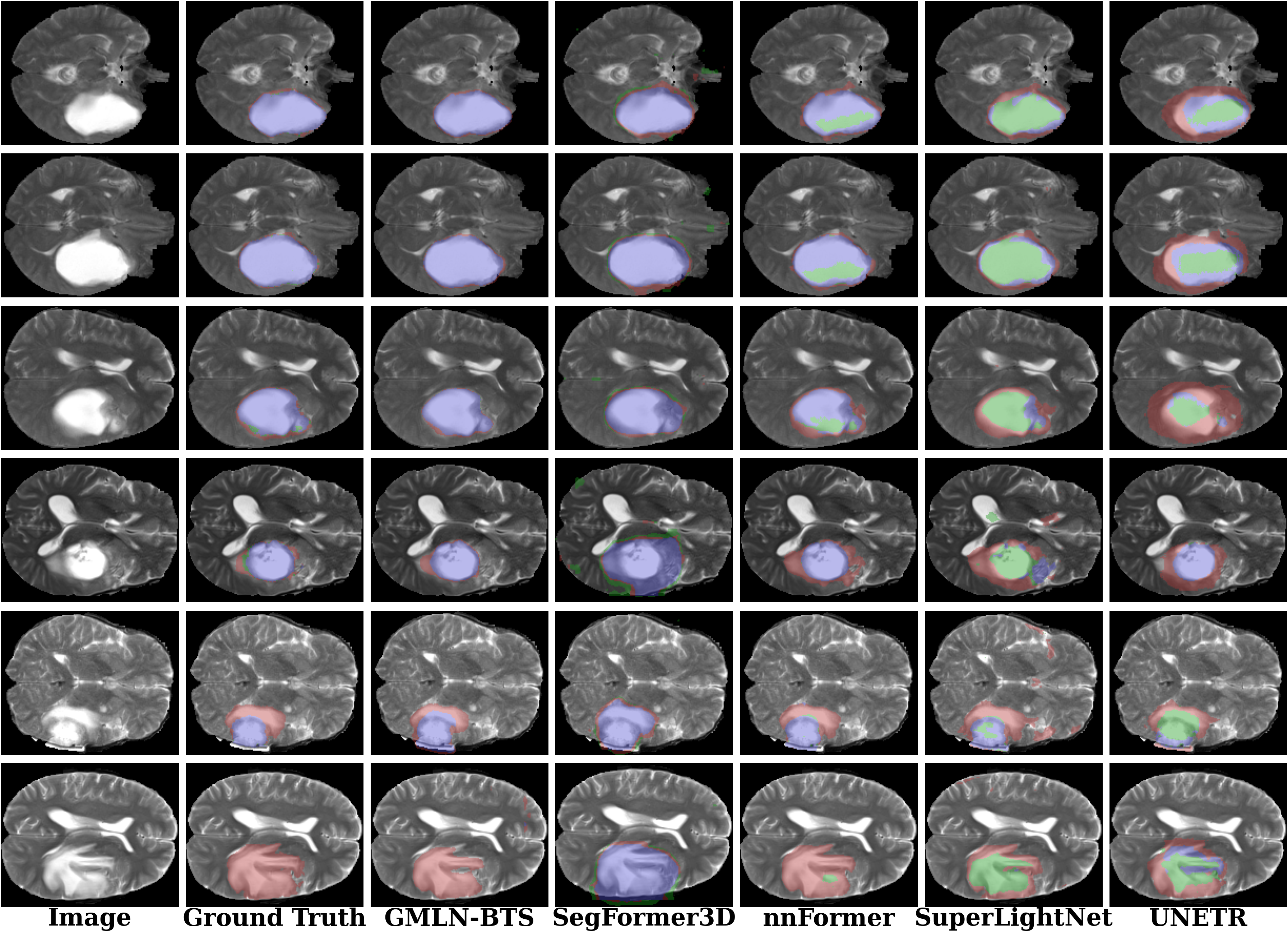}
    \vspace{-10pt}
    \caption{Qualitative visualization results of different models on the BraTS2017 dataset.}
    \label{fig:visual}
    \vspace{-10pt}
\end{figure}

\noindent\textbf{Ablation Study.} Table \ref{tab:ablation} illustrates the performance gains attributed to the G2MCIM (2.3\%), M2AE (0.5\%), and VRUM (0.4\%) modules, validating their individual contributions. Qualitative results on the BraTS2017 dataset are visualized in Figure \ref{fig:visual}.

\section{Conclusion}
We present GMLN-BTS, a lightweight graph-based framework for efficient multi-modal brain tumor segmentation. Our architecture integrates three core components: a Modality-Aware Adaptive Encoder (M2AE) for robust feature extraction, a Graph-based Multi-modal Collaborative Interaction Module (G2MCIM) to explicitly model cross-modal dependencies, and a Voxel Refinement UpSampling Module (VRUM) for fine-grained boundary preservation. Evaluations on BraTS 2017, 2019, and 2021 benchmarks demonstrate that GMLN-BTS achieves state-of-the-art performance among lightweight models. With only 4.58M parameters, our method significantly reduces computational overhead while maintaining accuracy competitive with heavy-duty 3D Transformers. While current graph construction utilizes predefined relationships, future work will investigate dynamic graph learning to enhance clinical adaptability.

\newpage
\bibliographystyle{splncs04}
\bibliography{iclr2026_conference}

\end{document}

%% file: math_commands.tex
\usepackage{amsmath,amsfonts,bm}

\def\eqref#1{equation~\ref{#1}}

\def\1{\bm{1}}

\DeclareMathAlphabet{\mathsfit}{\encodingdefault}{\sfdefault}{m}{sl}
\SetMathAlphabet{\mathsfit}{bold}{\encodingdefault}{\sfdefault}{bx}{n}

